\documentstyle[preprint,aps,floats]{revtex}

 \input amssym.def   
 \input amssym.tex

\font\twelvemsb=msbm10 at 12pt
\font\ninemsb=msbm7 at 9pt
\font\sixmsb=msbm5 at 6pt
\newfam\Msbfam
\textfont\Msbfam=\twelvemsb
\scriptfont\Msbfam=\ninemsb
\scriptscriptfont\Msbfam=\sixmsb

\def\beq{\begin{equation}}
\def\eeq{\end{equation}}
\def\bi{\begin{itemize}}
\def\ei{\end{itemize}}
\def\beqar{\begin{eqnarray}}
\def\eeqar{\end{eqnarray}}

\newcommand{\rmd}{{\rm d\null}}

\def\lra{\mathop{\hbox to .5in{\rightarrowfill}}}

\let\varkappa\kappa

\skip\footins = 14pt % was 12
\begin{document}

%% \nonfrenchspacing
%% \flushbottom

\title{Anyon spin and the exotic central extension of the \\ 
planar Galilei group\footnotemark[1]}

\author{R. Jackiw and V.P. Nair\footnotemark[2]}

\footnotetext[1] {\baselineskip=12pt This work is supported
in part by funds provided by  the U.S.~Department of Energy
(D.O.E.) under contract
\#DE-FC02-94ER40818 and by NSF grant PHY-9605216.
\hfill MIT-CTP-2960, 
hep-th/0003130,  March 2000}

\footnotetext[2]{\baselineskip=12pt Permanent address: Physics Department, City College
of the CUNY, New York, NY 10031. }

\address{Center for Theoretical Physics\\ Massachusetts
Institute of Technology\\ Cambridge, MA ~02139--4307,
USA}

\maketitle
\begin{abstract}

We show that the second central extension of the Galilei group in $(2+1)$ dimensions
corresponds to spin, which can be any real number. 

\end{abstract}
\vskip .1in

It has been known for a long time that the Galilei group in
$(2+1)$ dimensions admits a two-parameter central
extension \cite{LLB}. One of these, which is present in all dimensions,
is well known to be the mass $m$.
The second extension has remained somewhat more exotic.
The classical dynamics of one-particle representations of the
group can
be discussed via the symplectic form on the corresponding
coadjoint orbit. With the extension parameters $m, \kappa$ the
Cartan-Poincar\'e form is
\begin{equation}
\omega=\rmd p^i \rmd x^i + { \kappa\over 2m^2} \epsilon^{ij}
\rmd p^i
\rmd p^j -{p^i\rmd p^i~ \rmd t\over m}
\label{eq:1}
\end{equation}
The first two terms on the right give the symplectic
two-form. We include the last term $-dH\rmd t$ so that classical
evolution is given by the kernel of $\omega$.  Recently
there have been some discussions of the second extension
and of the symplectic form (\ref{eq:1}) \cite{Luk,duval}.

It has also been known for some time that the Poincar\'e
group in $(2+1)$ dimensions admits unitary representations
of arbitrary spin \cite{bin,JN}. Spin is a free parameter which does not
have to be quantized.  The symplectic structure for a free
relativistic particle can be taken to be 
\begin{equation}
\Omega=-\rmd p_a \rmd x^a + {sc^2\over 2} \frac{\epsilon^{abc}
p_a \rmd p_b \rmd p_c}{(p^2)^{\frac{3}{2}}}
\label{eq:2}
\end{equation}
Here $s$ is the spin of the particle. The constraint
$p^2-m^2c^2=0$ can be imposed consistently to generate the
dynamics. (Our metric tensor is $g_{ab}=diag(1,-1,-1)$.)

The present comment is to the effect that the nonrelativistic
limit of (\ref{eq:2}) gives (\ref{eq:1}), thereby identifying
the second extension parameter $\kappa$ as the spin of the
particle.

This reduction is completely straightforward. We solve
$p^2-m^2c^2=0$ to write $p_0=\sqrt{p^ip^i+m^2c^2}$ and do a nonrelativistic
$\frac{1}{c}$-expansion to get  $p_0 \approx
mc+\frac{p^i p^i}{2mc}$.  The leading term in (\ref{eq:2})
then becomes
\begin{equation}
\Omega\approx -
\frac{p^i\rmd p^i\rmd t}{m}+ \rmd p^i\rmd
x^i+{s\over 2} 
\frac{\epsilon^{ij}\rmd p^i\rmd p^j}{m^2}+ \cdots
\label{eq:3}
\end{equation}
which agrees with (\ref{eq:1}), with the identification
$s=\kappa$.  The nonrelativisitic limit of
(\ref{eq:2}) thus gives the extended Galilei group and the
second extension parameter corresponds to spin.
The second extension of the Galilei group may also be
characterized by the modified commutation rule for the
Galilean boosts $K_i$, viz.,
\begin{equation}
[K_i,K_j]=i\kappa\epsilon_{ij}
\label{eq:4}
\end{equation}

In the relativistic case, the generator of $SO(2,1)$ rotations
is identified as 
\begin{equation}
J_a=\epsilon_{abc}x^bp^c+\frac{sc^2p_a}{\sqrt{p^2}}
\label{eq:5}
\end{equation}
 These obey the $SO(2,1)$ algebra
\begin{equation}
[J^a, J^b]=i\epsilon^{abc} J_c
\label{eq:6}
\end{equation}
The boost generators are defined by $J^i=c\epsilon^{ij}K_j$
and obey $[K_i,K_j]=i\epsilon_{ij}(J_0/c^2).$\ This simplifies to
(\ref{eq:4}) when we use $p_0\approx mc+{p^ip^i\over 2mc}$,
to leading order in $\frac{1}{c}$. \ (The last term in
(\ref{eq:5}) dominates 
$J_0$.)  The expressions (\ref{eq:5}) can also be similarly
reduced and agree with the moment maps 
(or generators of symmetry transformations) for (\ref{eq:1}), as
evaluated for instance in \cite{Luk,duval}, up to additive constants.
The canonical commutation rules which follow from
(\ref{eq:2}) are the standard ones except for
\begin{equation}
[x^a, x^b]=isc^2~\frac{ \epsilon^{abc} p_c}{(p^2)^\frac{3}{2}}
\label{eq:7}
\end{equation}
In the nonrelativistic limit,
\begin{equation}
[x_i,x_j]\approx\frac{is\epsilon_{ij}}{m^2},
\label{eq:8}
\end{equation}
a peculiarity
which has been noticed in the case of (\ref{eq:1}) \cite{Luk,duval}. The noncommutativity
of the coordinates may seem strange, but from the
point of view of a spinning particle, 
it is not so exotic. It is well known that in relativistic
theories the true
position operator is not the naive one and one has to make a Foldy-Wouthuysen
type transformation to obtain it.
The transformation of (\ref{eq:7}) to a canonical set of variables 
involves the use of the Dirac monopole potential in $p$-space \cite{JN,CNP}.
For the nonrelativistic case of (\ref{eq:8}) the corresponding redefinition
is
\begin{equation}
x^i = q^i - {s\over 2m^2} \epsilon^{ij} p^j
\label{eq:9}
\end{equation}
where $q^i,p^i$ are canonically conjugate.

The coupling of the relativistic particle to electromagnetic
fields has been done via the addition of
$eF=\frac{1}{2}eF_{ab}\rmd x^{a} \rmd x^b$ to (\ref{eq:2}) \cite{CNP}.
Among
other results, this showed that the gyromagnetic ratio is 2
for particles in $(2+1)$ dimensions for any value of spin.
Recently, the coupling of the Galilean particle to
electromagnetism has been analyzed by adding $eF$ to
(\ref{eq:1}) \cite{duval}. Needless to say, the results agree with the
corresponding nonrelativistic limit of reference \cite{CNP}.

Finally, note that in the ultrarelativistic limit $c\rightarrow 0$,
the anomalous spin contribution to (\ref{eq:2}) vanishes. Since this limit also
describes massless particles, we regain the result that planar massless
particles carry no spin \cite{DJ}.

\end{document}